\documentclass[aps,prl,twocolumn,unsortedaddress,superscriptaddress,showpacs]{revtex4-1}

\usepackage{amsmath}
\usepackage{hyperref}
\usepackage{graphicx}
\usepackage{physics}
\usepackage{isotope}
\usepackage[capitalize]{cleveref}

\hypersetup{colorlinks=true,citecolor=[rgb]{0,0,0.8},
linkcolor=[rgb]{0,0,0.8},urlcolor=[rgb]{0,0,0.8}}

\begin{document}
\title{Reply to Comment on ``Is a Trineutron Resonance Lower in Energy than a Tetraneutron
Resonance?'' by A.~Deltuva and R.~Lazauskas}

\author{S.~Gandolfi}
\email{stefano@lanl.gov}
\affiliation{Theoretical Division, Los Alamos National Laboratory,
Los Alamos, New Mexico 87545, USA}
\author{H.-W.~Hammer}
\email{hans-werner.hammer@physik.tu-darmstadt.de}
\affiliation{Institut f\"ur Kernphysik, 
Technische Universit\"at Darmstadt, 64289 Darmstadt, Germany}
\affiliation{ExtreMe Matter Institute EMMI, 
GSI Helmholtzzentrum f\"ur Schwerionenforschung GmbH, 64291 Darmstadt,
Germany}
\author{P.~Klos}
\email{pklos@theorie.ikp.physik.tu-darmstadt.de}
\affiliation{Institut f\"ur Kernphysik, 
Technische Universit\"at Darmstadt, 64289 Darmstadt, Germany}
\affiliation{ExtreMe Matter Institute EMMI, 
GSI Helmholtzzentrum f\"ur Schwerionenforschung GmbH, 64291 Darmstadt,
Germany}
\author{J.~E.~Lynn}
\email{joel.lynn@physik.tu-darmstadt.de}
\affiliation{Institut f\"ur Kernphysik, 
Technische Universit\"at Darmstadt, 64289 Darmstadt, Germany}
\affiliation{ExtreMe Matter Institute EMMI, 
GSI Helmholtzzentrum f\"ur Schwerionenforschung GmbH, 64291 Darmstadt,
Germany}
\author{A.~Schwenk}
\email{schwenk@physik.tu-darmstadt.de}
\affiliation{Institut f\"ur Kernphysik, 
Technische Universit\"at Darmstadt, 64289 Darmstadt, Germany}
\affiliation{ExtreMe Matter Institute EMMI, 
GSI Helmholtzzentrum f\"ur Schwerionenforschung GmbH, 64291 Darmstadt,
Germany}
\affiliation{Max-Planck-Institut f\"ur Kernphysik, Saupfercheckweg 1, 
69117 Heidelberg, Germany}

\maketitle

In Ref.~\cite{Gandolfi:2016bth} we presented calculations for
three- and four-neutron ($3n$ and $4n$) states in the
presence of an external trapping potential. These calculations were
extrapolated to the limit of zero trap depth, as in
Ref.~\cite{pieper2003},
and showed the remarkable feature that these extrapolations point
to a common positive energy scale, independent of the trap geometries
considered. Based on calculations for a two-body resonance, where the
same extrapolation procedure correctly locates the resonance energy,
we suggested that our results support the possible
observation of a tetraneutron resonance~\cite{Kisamori:2016jie}, and
provide indications that a $3n$ resonance might also exist at an
energy below a possible $4n$ resonance.
We did not claim that a $3n$ or $4n$ resonance definitely
exists, nor did we quantify their widths.

The question of few-neutron resonances is an
interesting and challenging problem with many conflicting theoretical
results at present~\cite{bertulani2002,timofeyuk2003,pieper2003,
Lazauskas:2005,Hiyama:2016,shirokov2016,Gandolfi:2016bth,Fossez:2016dch,Greene:2018}
including Ref.~\cite{Lazauskas:2005fy}, which we regrettably missed in our
Letter and the more recent Ref.~\cite{Deltuva:2018xoa}, which
already put forward the arguments raised in the Comment by Deltuva
and Lazauskas~\cite{Deltuva:2019ngx}.

\begin{figure}[t!]
\begin{center}
\includegraphics[width=0.9\columnwidth]{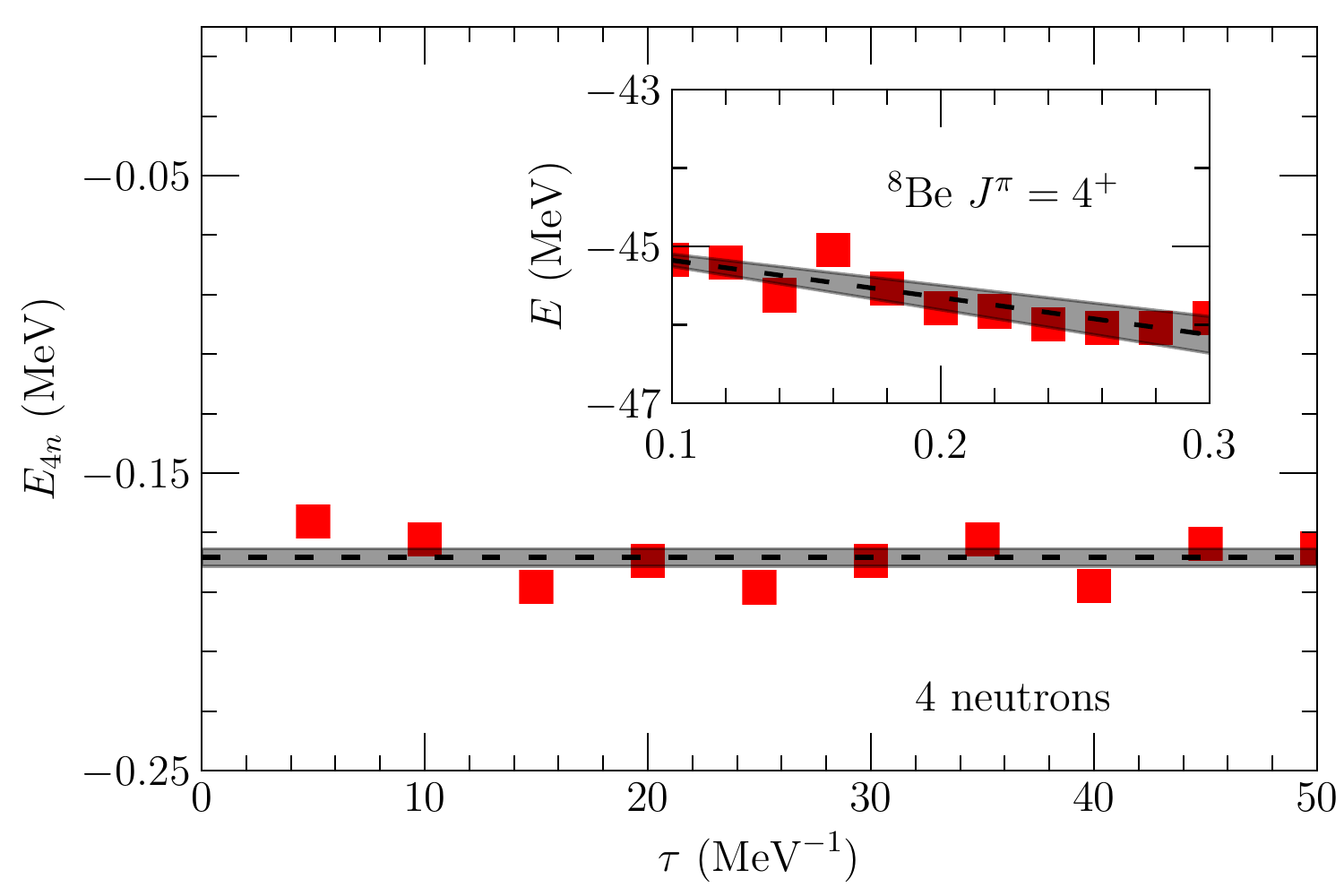}
\end{center}
\caption{The 4$n$ energy in an external Woods-Saxon well with
   $V_0=-1.25$~MeV and $R_\text{WS}=6.0$~fm as a function of
   imaginary time for the local chiral N$^2$LO interactions used in 
   Ref.~\cite{Gandolfi:2016bth}.
   The inset (extracted from Ref.~\cite{Pastore:2014oda})
   shows that an unbound $J^\pi=4^+$ state of \isotope[8]{Be} 
   decays very rapidly as a function imaginary time evolution (note
   the axes are in the same units).}
\label{fig:evstau2}
\end{figure}

The arguments presented in the Comment largely rely on ideas related to
the analytic continuation in the coupling constant (ACCC) method, where
the Hamiltonian of the system is written as
$H(\lambda)=H_0+\lambda H_\text{att}$, with $H_\text{att}$ the attractive
part of the $(\lambda=1)$ original Hamiltonian.
However, we point out that ACCC is not the same as applying
an external trap, which is the procedure we employ in our Letter.
Therefore, while we agree that bound dineutrons emerge
early on as the trap depth $V_0$ is nonzero, we do
not agree with the authors' conclusion that ``in the presumed $E_{4n}\approx0$
region \ldots [t]he tetraneutron states \ldots are not
true bound states.''
It is not clear what the authors mean by ``true bound states.''
Bound states are states whose wave functions have compact
support. This is the case for all of our calculated $3n$ and $4n$ states
in the trap.
In our Letter, we used the auxiliary field diffusion Monte Carlo method,
which converges to the lowest energy eigenstate
with the relevant quantum numbers of a given Hamiltonian.
There are cases where diffusion Monte Carlo methods have been applied to
states that decay.
For example, the unbound nucleus \isotope[8]{Be} was calculated
using the Green's function Monte Carlo method in Ref.~\cite{Pastore:2014oda}.
In this case, the states decay asymptotically to two $\alpha$ clusters, and
this decay is observed clearly, e.g., in the evolution of the $4^+$ energy
even after a short imaginary time, as shown in the inset of \cref{fig:evstau2}.
As \cref{fig:evstau2} also shows, for the 4$n$ system in the region in
question, where $E_{2n}<E_{4n}$ (e.g., Woods-Saxon (WS) well depth
$V_0=-1.25$~MeV and WS radius $R_\text{WS}=6.0$~fm), we observe no such
decay in the energy over a very long imaginary time evolution.
This suggests that this $4n$ state is more complex than a
pair of dineutrons, or a dineutron with a pair of neutrons.
Moreover, we have checked that including in the extrapolation only the
points where $E_{4n}<E_{2n}$ for $R_\text{WS}=7.5$~fm still identifies
the potential $4n$ resonance at approximately 2.5~MeV.

\begin{figure}[t!]
\begin{center}
\includegraphics[width=1.0\columnwidth]{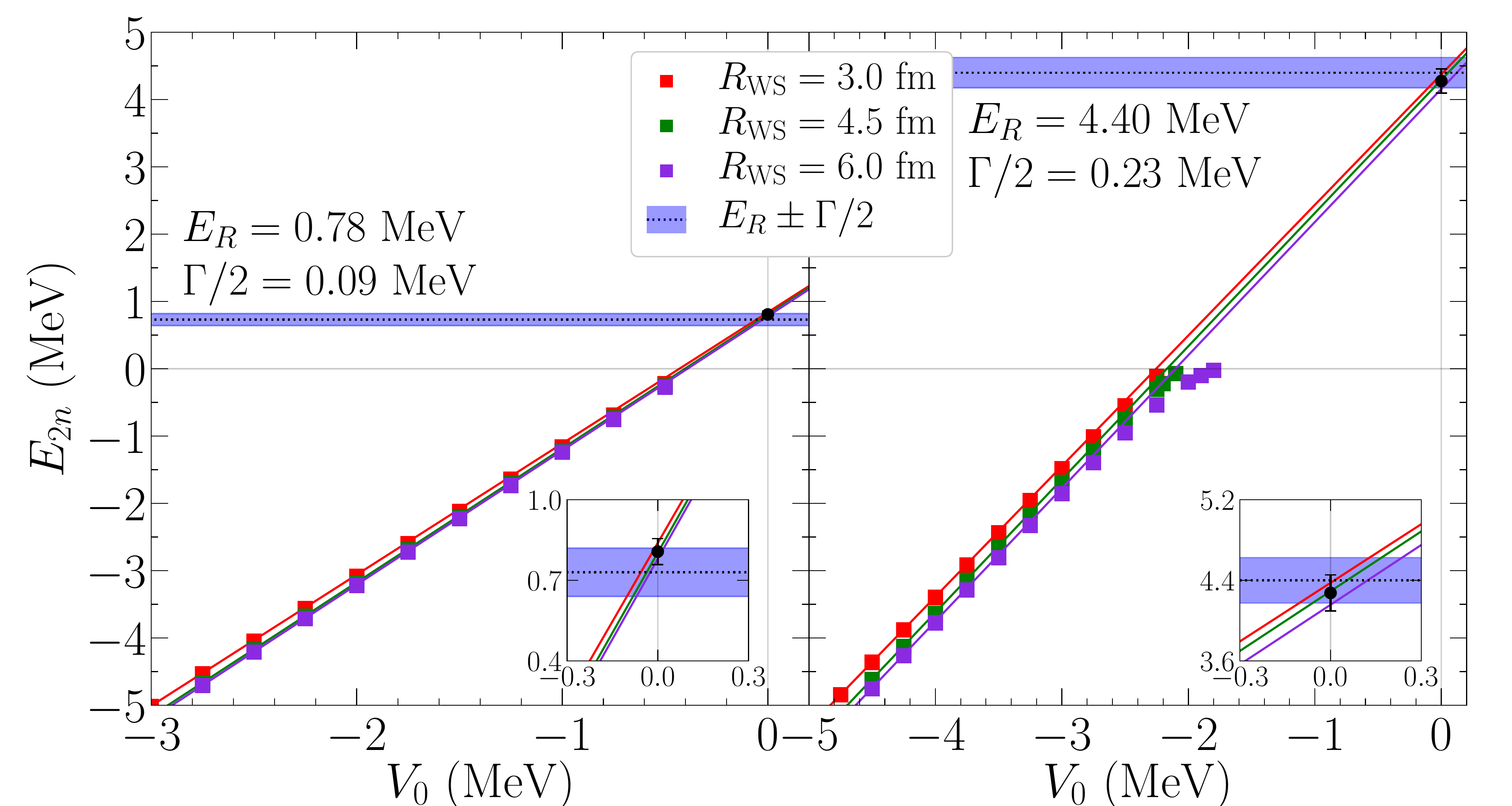}
\end{center}
\caption{The energy of two neutrons trapped in various Woods-Saxon wells
interacting via an $S$-wave (two-Gaussian) potential as in Ref.~\cite{Gandolfi:2016bth}
tuned to give two different resonances.
The linear extrapolations to zero well depth correctly give the
position of the resonance in both cases.}
\label{fig:both_new_resonances}
\end{figure}

Regarding the extrapolation procedure itself, we are aware that some
care is needed, which is why we sought to establish that our
extrapolation works well in the two-body $S$-wave (two-Gaussian)
potential case as discussed in Ref.~\cite{Gandolfi:2016bth}.
To reinforce this point, we have performed additional calculations for
two two-body resonances shown in~\cref{fig:both_new_resonances}.
As can be seen in~\cref{fig:both_new_resonances}, our
extrapolation procedure correctly identifies the locations of the two
resonances within the uncertainties of the fit.
Furthermore, as our Letter notes, using the current extrapolation, we
cannot make a comment about the width.
It is entirely possible that the width is very broad (see
also Ref.~\cite{Fossez:2016dch}) and therefore the resonance would have
little or no effect on the observable scattering dynamics.
In fact, we acknowledge that our current extrapolation cannot
distinguish between a resonance and a virtual state.

In conclusion, the existence of few-neutron structures is ultimately a 
question that experiments have to decide. It remains an intriguing open question
whether these systems exhibit resonances, virtual states, or other
localized features of the cross section unrelated to $S$-matrix poles.

We thank S.~Dietz for useful discussions and benchmark calculations.
This work was supported by the U.S. Department of Energy under Contract
No.~DE-AC52-06NA25396, the NUCLEI SciDAC project, the ERC Grant
No.~307986 STRONGINT, and the Deutsche Forschungsgemeinschaft through
Grant No.~SFB~1245.

\bibliography{reply_comment}

\end{document}